\documentclass[aps,pra,preprintnumbers,groupedaddress,nofootinbib,showpacs,twocolumn]{revtex4}

\usepackage{graphics}
\usepackage{picins}
\usepackage[colorlinks=true]{hyperref}
\usepackage{color}
\usepackage{graphicx}
\begin{document}

\title{ Relativistic  Hydrogen-Like Atom on a
Noncommutative Phase Space}

\author{Huseyin Masum$^a$}

\author{Sayipjamal Dulat$^a$}
\email{dulat98@yahoo.com}

\author{Mutallip  Tohti$^b$}
\affiliation{\it  $^a$ School of Physics Science and Technology, Xinjiang
University, Urumqi, 830046, China}

\affiliation{\it  $^b$Radio Therapy center, Xinjiang Tumor Hospital, Urumqi, 830000, China}




\begin{abstract}
The energy levels of hydrogen-like atom on a noncommutative phase space were studied in the framework of relativistic quantum mechanics. The leading order corrections to energy levels $2S_{1/2}$, $2P_{1/2}$ and $2P_{3/2}$ were obtained  by using the  $\theta$ and the $\bar\theta$ modified Dirac Hamiltonian of hydrogen-like atom on a noncommutative phase space.  The degeneracy of the  energy levels $2P_{1/2}$ and $2P_{3/2}$ were removed completely by $\theta$-correction. And  the $\bar\theta$-correction shifts these energy levels.
\end{abstract}

\maketitle

\section{Introduction}
The approach to noncommutative quantum field theory based on star
products and Seiberg-Witten maps allows for the generalization of
the standard model of particle physics to the case of
noncommutative space-time. Since noncommutative quantum field
theory may solve the puzzles of the standard model, there are many papers concerning the quantum field theory on a noncommutative
space-time \cite{NCQEDGG}-\cite{NCQED}. Apart from these studies,
much research has been devoted to the study of various aspects of
quantum mechanics (QM) on a noncommutative space (NCS) and a
noncommutative phase space(NCPS), because the main goal of
noncommutative quantum mechanics (NCQM)
is to find measurable effects of noncommutativity. For
example, the papers \cite{10}-\cite{12} was devoted to study the
Aharonov-Bohm phase on a NCS and a NCPS. A lower bound
$1/\sqrt\theta \geq 10^{-6}GeV$ for the space noncommutativity
parameter was obtained \cite{10}. The Aharonov-Casher phase for a
spin-1/2 and a spin-1 particle on a NCS and a NCPS has been studied
in Refs. \cite{13}-\cite{16}, and a limit $1/\sqrt\theta \geq
10^{-7}GeV$ for the space noncommutativity parameter was otained
\cite{13}. The noncommutative quantum Hall effect has been studied
in Refs. \cite{17}-\cite{19}, and the authors of Ref. \cite{17}
found a lower limit of $1/\sqrt\theta \geq 10GeV$ on the
noncommutativity parameter.  Ref. \cite{20,21} discussed the
noncommutative spin Hall effect (SHE), and obtained interesting results. Furthermore,
 a lower limit of $1/\sqrt\theta \geq 10^{-12}
GeV$ on the noncommutative parameter  was given in Ref.
\cite{21}.
The authors in Refs. \cite{HAncs,HAncps} studied
hydrogen atom spectrum in the nonrelativistic quantum mechanics
framework both on a NCS and a NCPS, respectively, and the authors in Ref.\cite{HAncs} found the
constraint on $\theta$ is $1/\sqrt\theta \geq 10^{4}GeV$.
Reference~\cite{Haghighat} provided the
constraint: $1/\sqrt\theta \geq 3GeV$ by studying the transitions in the helium atom.  A possibility of testing spatial noncommutativity
via cold Rydberg atoms is suggested in Ref.\cite{NCPR}.  In order to refine the work in Ref. \cite{HAncs}, the authors in Ref. \cite{RHAncs} has studied the hydrogen atom on a NCS in the framework of the $\theta$ modified Dirac equation with Coulomb potential, and they showed that the degeneracy of the energy levels $2P_{1/2}$ and $2P_{3/2}$ were  removed completely. 

In this paper we study the NCPS effects on the energy levels of relativistic hydrogen-like atom. To begin, we must define what we mean by ``noncommutative phase space". NCPS is a deformation of ordinary space in
which the space and momentum coordinate operators satisfy the
following relation($\hbar = c = 1$):
\begin{eqnarray}\label{non cumu r}
\big[\hat{x}_{i},\hat{x}_{j}\big]=i\mathbf{\Theta}_{ij}\;,\hspace{0.2cm}
\big[\hat{p_{i}},\hat{p_{j}}\big]=i\mathbf{\bar{\Theta}}_{ij}\;,\hspace{0.2cm}
\big[\hat{{x}_{i}},\hat{p_{j}}\big]=i\delta_{ij}\;,
\end{eqnarray}
where $\mathbf{\Theta}_{ij}$ and $\mathbf{\bar{\Theta}}_{ij}$ are  the totally antisymmetric real tensors on a NCPS; $\hat{x}$, $\hat{p}$ are the coordinate and momentum operators on a NCPS.
In three dimensional NCPS ($i,j=1,2,3$), we can define a vector
$\theta=(\theta_{1},\theta_{2},\theta_{3})$ and $\bar\theta=(\bar{\theta}_1,\bar{\theta}_2,\bar{\theta}_3)$  with $\theta_{i}$
 and $\bar{\theta}_i$ satisfy $\mathbf\Theta_{ij}=\epsilon_{ijk}\theta_k$, $\mathbf{\bar{\Theta}}_{ij}=\epsilon_{ijk}\bar{\theta}_k$, here $\epsilon_{ijk} $ is the Levi-Civita symbol.

On a NCPS, the normal product of two arbitrary functions should be replaced by the star product(Moyal-Weyl product). For example, the time independent Schr\"{o}dinger equation on a NCPS is
\begin{equation}\label{NC-shronigger}
H(\emph{p}, \emph{r})*\Psi(\emph{r})=E\Psi(\emph{r}).
\end{equation}
Here $H(x,p)$ is the usual Hamiltonian operator.
On a NCPS star product between two functions is defined as
\begin{eqnarray}
(f*g)(x,p)&=&e^{\frac{i}{2\alpha^{2}}\Theta_{ij}\partial^{x}_{i}\partial^{x}_{j}
+\frac{i}{2\alpha^{2}}\bar{\Theta}_{ij}\partial^{p}_{i}\partial^{p}_{j}}f(x,p)g(x,p)
\nonumber\\
&=&f(x,p)g(x,p) + \frac{i}{2\alpha^{2}}\Theta_{ij}\partial^{x}_{i}f\partial^{x}_{j}g|_{x_{i}=x_{j}}\nonumber\\
&+&\frac{i}{2\alpha^{2}}\bar{\Theta}_{ij}\partial^{p}_{i}f\partial^{p}_{j}g|_{p_{i}=p_{j}}
+O(\theta^{2},\bar{\theta}^{2})\;,
\end{eqnarray}
here $f(x,p)$ and $g(x,p)$ are two arbitrary functions, the scaling
constant $\alpha$ is related to the noncommutativity parameters $\theta$
and $\bar{\theta}$ via $\theta\bar{\theta}=4\alpha^{2}(1-\alpha^{2})$.
To replace the star product in Schr\"{o}dinger
Eq.(\ref{NC-shronigger}) with a usual product, first we need to
replace $x_{i}$ and $p_{i}$ with a generalized Bopp's shift as
 \begin{eqnarray}\label{ncps-shift}
 \hat{x}_{i}&=& \alpha x_{i}-\frac{1}{2\hbar \alpha}\Theta_{ij}p_{j}\;,\;\nonumber\\
 \hat{p}_{i}&=&\alpha p_{i}+\frac{1}{2\hbar \alpha}\bar{\Theta}_{ij}x_{j}\;, \;\;\; i,j =
 1,2,...,n\;.
\end{eqnarray}
Thus on a NCPS the Schr\"{o}dinger Eq.(\ref{NC-shronigger})
becomes,
 \begin{equation}
 H(\hat{\emph{p}},\hat{\emph{r}})\Psi(\emph{r})= E\Psi(\emph{r}),\hspace{0.2cm}
 H(\emph{p},\emph{r})*\Psi(\emph{r}) \equiv H(\hat{\emph{p}},\hat {\emph{r}})\Psi(\emph{r}),
\end{equation}
here $\emph{r}$ and $\emph{p}$ are the space and the momentum operators on an ordinary space.

This paper is organized as follows. In section II, first we give usual Dirac Hamiltonian for an electron  in the Coulomb field, and we list corresponding eigenfunctions and eigenvalues. Then we provide the NCPS Dirac Hamiltonian for the hydrogen-like atom. In Section III, using the perturbation theory, leading order correction to the energy levels due to space-space and momentum-momentum noncommutativity  are obtained. Conclusions are given in the last section.

\section{Dirac Hamiltonian with Coulomb Field on a NCPS}
Before writing Dirac Hamiltonian with Coulomb field on a NCPS,
 first, we list usual Dirac Hamiltonian, Dirac equation,  energy spectrum  and eigenfunctions by following Refs.\cite{Greiner}-\cite{Voronov}. Then we provide the $\theta$ and $\bar\theta$ modified Dirac Hamiltonian  on a NCPS.
On an ordinary space, the Dirac Hamiltonian for an electron with charge $-e(e>0)$
and mass $m$ in the Coulomb field of a nucleus $Ze$ is given by
\begin{eqnarray}\label{Homiltonian on CS}
H=\tilde{\alpha} \cdot \mathbf{p} + m\gamma^0 +V(r) ,
\end{eqnarray}
where $\tilde{\alpha}_i=\gamma_0\gamma_i$, $\gamma_i$ are the Dirac matrices, $\mathbf{p}=-i\nabla$, $V(r)=-Ze^{2}/r$.
The stationary Dirac equation for  electron is
\begin{eqnarray}
&&H\Psi_{n,j,l,j_z} = E_{n,j}\Psi_{n,j,l,j_z}\;,
\end{eqnarray}
where energy eigenvalues of electron in atom with a Coulomb potential is
\begin{eqnarray}\label{E zero}
E_{n,j}&=&\frac{m}{\sqrt{1+\frac{(Ze^{2})^2}{\big[n-j-1/2+\sqrt{(j+1/2)^{2}-(Ze^{2})^2}\big]^2}}}\nonumber\\
&=&\frac{m}{\sqrt{1+(\frac{Ze^{2}}{\gamma_{n^{\prime}}+n^{\prime}})^{2}}}\;,
\end{eqnarray}
with
\begin{eqnarray}
\gamma_{n^\prime}&=&\sqrt{\kappa^2-(Ze^2)^2}\;,\hspace{0.5cm} n^{\prime}=n-|\kappa|=n-j-\frac{1}{2}\;,\nonumber\\
\kappa&=&\mp(j+\frac{1}{2})=:\bigg\{
\begin{array}{c}
-(j+\frac{1}{2}) \hspace{0.4cm} \textrm{for} \hspace{0.5cm} j=l+\frac{1}{2}\;,
\\
\\(j+\frac{1}{2}) \hspace{0.5cm} \textrm{for} \hspace{0.5cm} j=l-\frac{1}{2}\;,
\end{array}
\end{eqnarray}
 here $n=1,2,3,\cdots$ is the principal quantum number. From equation (\ref{E zero}) one can see that the energy eigenvalues only depend on the $n$, $j$ and  $Z$. The  wave functions $\Psi_{n,j,l,j_z}$ are also
eigenfunction of $\hat J^2$ and $\hat J_z$
\begin{equation}
\hat{J}^{2}\Psi_{n,j,l,j_z}=j(j+1)\Psi_{n,j,l,j_z},\hspace{0.4cm}
\hat{J}_{z}\Psi_{n,j,l,j_z}=j_z\Psi_{n,j,l,j_z},
\end{equation}
The corresponding wave functions $\Psi_{n,j,l,j_z}$ are given by
\begin{eqnarray}\label{bispinors}
\Psi_{n,j,l,j_z}&=&\left(
\begin{array}{c}
\varphi_{n,j,l,j_z}(r)
\\\chi_{n,j,l^{\prime},j_z}(r)
\end{array}\right),\nonumber\\
\varphi_{n,j,l,j_z}(r)&=&ig(\textit r)\;\Omega_{j,l,j_z}\bigg(\frac{\textbf r}{r}\bigg),\\
\chi_{n,j,l^{\prime},j_z}(r)&=&f(r)\;\Omega_{j,l^{\prime},j_z}\bigg(\frac{\textbf r}{r}\bigg),\nonumber
\end{eqnarray}
where $l^{\prime}=2j-l$.
The spherical spinors $\Omega_{j,l,j_z}$ are eigenfunction of the operators $\hat L^2$, $\hat J^2$ and
$\hat S^2=(\frac{1}{2}\hat\sigma)^2$ with eigenvalues $l(l+1)\hbar^2$, $j(j+1)\hbar^2$ and $\frac{3}{4}\hbar^2$ respectively; the explicit forms of the $\Omega_{j,l,j_z}$ for the cases $j=l+\frac{1}{2}$ and $j=l-\frac{1}{2}(j\geq\frac{1}{2})$ are
\begin{eqnarray}\label{omega}
\Omega_{j,l,j_z}&=&\left(
\begin{array}{c}
\sqrt{\frac{j+j_z}{2j}}\;Y_{l,j_z-\frac{1}{2}}(\vartheta,\varphi)
\\\sqrt{\frac{j-j_z}{2j}}\;Y_{l,j_z+\frac{1}{2}}(\vartheta,\varphi)
\end{array}\right)\hspace{0.7cm} \textrm{for}\hspace{0.1cm} j=l+\frac{1}{2},\nonumber\\
\Omega_{j,l,j_z}&=&\left(
\begin{array}{c}
-\sqrt{\frac{j-j_z+1}{2j+2}}\;Y_{l,j_z-\frac{1}{2}}(\vartheta,\varphi)
\\\sqrt{\frac{j+j_z+1}{2j+2}}\;Y_{l,j_z+\frac{1}{2}}(\vartheta,\varphi)
\end{array}\right)\hspace{0.1cm} \textrm{for}\hspace{0.1cm} j=l-\frac{1}{2},\nonumber\\
\end{eqnarray}
here the root factors of (\ref{omega}) are Clebsch-Gordon coefficients, $Y_{A,B}(\vartheta,\varphi)$ are the spherical harmonics; the normalized radial wave functions $f(r)$ and $g(r)$ in (\ref{bispinors}) are
\begin{eqnarray}
\begin{array}{c}
g(r)
\\f(r)
\end{array}\bigg\}&=&\frac{\pm(2\lambda_{n^{\prime}})^{3/2}}{2\Gamma(\beta_{n^{\prime}})} \sqrt{\frac{\lambda_{n^{\prime}}^{\pm}\Gamma(n^{\prime}+\beta_{n^{\prime}})} {\eta_{n^{\prime}}(\eta_{n^{\prime}}-\kappa)n^{\prime}!}}(2 r \lambda_{n^{\prime}})^{\gamma_{n^{\prime}}-1}\nonumber\\
&&\times
e^{-r\lambda_{n^{\prime}}}\{(\eta_{n^{\prime}}-\kappa)
\Phi(-n^{\prime},\beta_{n^{\prime}};2r \lambda_{n^{\prime}})\nonumber\\
&&\hspace{1.2cm}\mp n^{\prime}\Phi(1-n^{\prime},\beta_{n^{\prime}};2r \lambda_{n^{\prime}})\}\;,
\end{eqnarray}
with
\begin{eqnarray}\label{def of eta gamma}
\lambda_{n^{\prime}}&=&\sqrt{m^2-E_{n,j}^2}\;,\hspace{0.8cm} \lambda^{\pm}_{n^{\prime}}=\sqrt{1\pm\frac{E_{n,j}}{m}}\;, \nonumber \\
\eta_{n^{\prime}}&=&\frac{(n^{\prime}+\gamma)m}{E_{n,j}}\;, \hspace{0.8cm} \beta_{n^{\prime}}=2\gamma+1\;,
\end{eqnarray}
and the hypergeometric confluent function  $\Phi(a,b;z)$ \cite{hypergeometric confluent and Gamma} is
\begin{equation}
\Phi(a,b;z)=1+\frac{a}{b}z+\frac{a(a+1)}{b(b+1)}\frac{z^2}{2!}+\dots \hspace{0.3cm}\;.
\end{equation}
As well as  the Gamma function \cite{hypergeometric confluent and Gamma} is
\begin{equation}
\Gamma(n)=\int^{\infty}_{0}t^{n-1}e^{-t}dt\;,\;\;\Gamma(n+1)=n\Gamma(n)\; .
\end{equation}

On a NCPS, the Dirac Hamiltonian (\ref{Homiltonian on CS}) can be written as
\begin{equation}\label{H}
\hat H(\hat p,\hat r)=\tilde{\alpha} \cdot \hat{\mathbf{p}}+ m\gamma^{0}-\hat V(\hat r),
\end{equation}
where the Coulomb potential with noncommutative correction terms is
\begin{eqnarray}\label{potential}
\hat V(\hat r)& = &-\frac{Ze^{2}}{\sqrt{\hat x_i \hat x_i}} \nonumber\\
&=& -\frac{Ze^{2}}{\sqrt{(\alpha x_{i}-\frac{1}{2\alpha}\Theta_{ij}p_{j})(\alpha
x_{i}-\frac{1}{2\alpha}\Theta_{ik}p_{k})}} \nonumber\\
& =& - \frac{Ze^{2}}{\alpha r} - \frac{Ze^{2}}{4\alpha^{3}r^{3}}(\mathbf{L}\cdot {\theta}) +
O(\theta^{2}).
\end{eqnarray}
By (\ref{ncps-shift}) and (\ref{potential}), we rewrite (\ref{H}) as
\begin{eqnarray}\label{E}
\hat H&=& \alpha \tilde{\alpha} \cdot \mathbf{p} + m\gamma^{0}  -
\frac{Ze^{2}}{\alpha r} - \frac{Ze^{2}}{4\alpha^{3}}\big(\frac{\mathbf{L}\cdot\theta}{r^{3}}\big)\nonumber\\
&& +\frac{1}{4\alpha}\tilde{\alpha}\cdot(\mathbf{r}\times\bar{\theta})+O(\theta^{2}) + O(\bar\theta^{2})\nonumber\\
&=&\alpha(\tilde{\alpha}\cdot \mathbf{p} + m^{\prime}\gamma^{0} -
\frac{Z{e^{\prime}}^{2}}{r}) - \frac{Z{e^{\prime}}^{2}}{4\alpha}\big(\frac{\mathbf{L}\cdot\theta}{{r}^{3}}\big)\nonumber\\
&&+\frac{1}{4\alpha}\tilde{\alpha}\cdot(\mathbf{r}\times\bar{\theta})+O(\theta^{2}) + O(\bar\theta^{2}) \nonumber\\
&=& H^{\prime} + H^{\theta} + H^{\bar{\theta}} +O(\theta^{2}) + O(\bar\theta^{2}),
\end{eqnarray}
with
\begin{eqnarray}\label{H prime H theta H bar theta}
H^{\prime}&=&\alpha(\tilde{\alpha}\cdot \mathbf{p} + m^{\prime}\gamma^{0} -
\frac{Z{e^{\prime}}^{2}}{r}),\hspace{0.3cm} H^{\theta}=-\frac{Z{e^{\prime}}^{2}}{4\alpha}\big(\frac{\mathbf{L}\cdot\theta}{{r}^{3}}\big),\nonumber\\ H^{\bar{\theta}}&=&\frac{1}{4\alpha}\tilde{\alpha}\cdot(\mathbf{r}\times\bar{\theta}) ,
\end{eqnarray}
where $e^\prime=\frac{\displaystyle e}{\displaystyle \alpha}$, $m^\prime=\frac{\displaystyle m}{\displaystyle \alpha}$; $\mathbf{L}$ is the orbital angular momentum operator.
Since the noncommutative corrections $H^{\theta}$ and $H^{\bar{\theta}}$ are very small compared to the $H$, the change in the hydrogen-like atom energy levels due to noncommutative parts $H^{\theta}$ and  $H^{\bar{\theta}}$ can always be treated  as some perturbation of the commutative counter part $H$. Up to first order in $\theta$ and $\bar\theta$,  one can use the usual wave functions in our forthcoming calculation of energy levels.

\section{Relativistic NCPS correction of energy levels}
By using the exact eigenfunctions of Dirac Hamiltonian $H$ and treating $H^{\theta}$ and $H^{\bar{\theta}}$
 as  perturbations, we can calculate the
modification of energy levels of the Hydrogen like atom on a NCPS in this section.
Thus NCS and NCPS effects on the energy levels are obtained by computing the eigenvalues of the secular matrix $E^{\theta}$ and $E^{\bar{\theta}}$, characterized by the average values of the operators $H^{\theta}$ and $H^{\bar{\theta}}$, with respect to the Dirac spinors $\Psi_{n,j,l,j_z}$, with angular momentum
selection rules, i.e., $\Delta j_z\equiv |j_z-j_z^\prime|=0,1$ and
$\Delta l \equiv|l-l^\prime|=0$. Thus matrix elements of $\hat H$ are
\begin{eqnarray}\label{cor E}
 E_{j_zj_z^{\prime}}(nL_j)&=&<\Psi_{n,j,l,j_z}\mid \hat H \mid\Psi_{n,j,l,j_z^{\prime}}> \nonumber\\
&=&<\Psi_{n,j,l,j_z}\mid H^\prime +H^\theta+H^{\bar{\theta}} \mid\Psi_{n,l,j,j_z^{\prime}}> \nonumber\\
&=&E_{n,j}^{\prime}+E^{\theta}_{j_zj_z^{\prime}}(nL_j)+E^{\bar{\theta}}_{j_zj_z^{\prime}}(nL_j)\;,
\end{eqnarray}
where $E_{n,j}^{\prime}$ is the eigenvalue of $H^\prime$, from equation (\ref{E zero}) and first equation of (\ref{H prime H theta H bar theta}), we have
\begin{eqnarray}\label{E}
E_{n,j}^{\prime}&=E_{n,j}+\Delta E_{n,j}\;,
\end{eqnarray}
with
\begin{eqnarray}\label{Delta E zero}
\Delta E_{n,j}&=&-2m\bigg[1+(\frac{Ze^{2}}{\gamma_{n^{\prime}}+n^{\prime}})^{2}\bigg]^{-\frac{3}{2}}\nonumber\\
&&\times\frac{Z^2e^3[(\gamma_{n^{\prime}}+n^{\prime})+Z^2e^4(\kappa^2-Z^2e^4)^{-\frac{1}{2}}]}{(\gamma_{n^{\prime}}+n^{\prime})^3}\nonumber\\
&&\times e(1-\alpha)\;.
\end{eqnarray}
Therefore the relativistic NCPS energy correction for a given energy level $nL_j$ is
\begin{equation}\label{Delta E}
\Delta  E_{j_z}(nL_j)=\Delta E_{n,j}+E^{\theta}_{j_z}(nL_j)+E^{\bar{\theta}}_{j_z}(nL_j)\;.\\
\end{equation}

 In leading order, the matrix elements of $H^{\theta}$ with respect to the Dirac spinors
$\Psi_{n,j,l,j_z}$ are defined as
\begin{eqnarray}\label{theta correction 1 }
E^{\theta}_{j_zj_z^{\prime}}(nL_j)&=&<\Psi_{n,j,l,j_z}\mid
H^{\theta}\mid\Psi_{n,j,l,j_z^{\prime}}>\nonumber\\
&=&-\frac{Ze^{2}}{4{\alpha}^3}\int^{\infty}_{0}\frac{dr}{r}[g(r)]^{\ast}\;g(r)\nonumber\\
&&\times \int^{4\pi}_{0}d\Omega\big[\Omega^{\dag}_{j,l,j_z}\;({\mathbf L} \cdot {\theta} )\;\Omega_{j,l,j_z^{\prime}}\big]\nonumber\\
&&-\frac{Ze^{2}}{4{\alpha}^3}\int^{\infty}_{0}\frac{dr}{r}[f(r)]^{\ast}\;f(r)\nonumber\\
&&\times \int^{4\pi}_{0}d\Omega\big[\Omega^{\dag}_{j,l^{\prime},j_z}\;({\mathbf L} \cdot {\theta} )\;\Omega_{j,l^{\prime},j_z^{\prime}}\big].\nonumber\\
\end{eqnarray}
Because $f(r)/g(r)\approx \upsilon/c$ ( $\upsilon$ is the velocity of electron in the first Bohr orbit ), so we can neglect the second term in (\ref{theta correction 1 }), then we have
\begin{eqnarray}\label{theta correction 2 }
E^{\theta}_{j_zj_z^{\prime}}(nL_j)&=&-\frac{Ze^{2}}{4{\alpha}^3}\int^{\infty}_{0}\frac{dr}{r}|g(r)|^2\nonumber\\
&&\times \int^{4\pi}_{0}d\Omega\big[\Omega^{\dag}_{j,l,j_z}\;({\mathbf L} \cdot {\theta} )\;\Omega_{j,l,j_z^{\prime}}\big]\nonumber\\
&=&-\frac{Ze^{2}}{4{\alpha}^3}\rho(nL_j)\Theta_{j_zj_z^{\prime}}(nL_j)\;,
\end{eqnarray}
with
\begin{eqnarray}\label{rho and theta}
\rho(nL_j)&=&\int^{\infty}_{0}\frac{dr}{r}|g(r)|^2,\nonumber\\
\Theta_{j_zj_z^{\prime}}(nL_j)&=&\int^{4\pi}_{0}d\Omega\big[\Omega^{\dag}_{j,l,j_z}\;({\mathbf L} \cdot {\theta} )\;\Omega_{j,l,j_z^{\prime}}\big]\label{thetamm}.
\end{eqnarray}
In leading order, matrix elements of  $H^{\bar \theta}$ with respect to the Dirac spinors $\Psi_{n,j,l,j_z}$ are defined as
\begin{eqnarray}\label{E bar theta}
E^{\bar{\theta}}_{j_zj_z^{\prime}}(nL_j)&=&<\Psi_{n,j,l,j_z}\mid
H^{\bar{\theta}}\mid\Psi_{n,j,l,j_z^{\prime}}>\nonumber\\
&=&<\Psi_{n,j,l,j_z}\mid
\frac{1}{4\alpha}\tilde{\alpha}\cdot(\mathbf r\times\bar{\theta})\mid\Psi_{n,j,l,j_z^{\prime}}>\nonumber\\
&=&-\frac{1}{4\alpha}\int^{\infty}_{0}drr^{3}[g(r)]^{\ast}\;f(r)\nonumber\\
&&\times \int^{4\pi}_{0}d\Omega\big[\Omega^{\dag}_{j,l,j_z}\;i{\sigma}\cdot(\mathbf n\times\bar{\theta})\;\Omega_{j,l^{\prime},j_z^{\prime}}\big]\nonumber\\
&&+\frac{1}{4\alpha}\int^{\infty}_{0}drr^{3}[f(r)]^{\ast}\;g(r)\nonumber\\
&&\times\int^{4\pi}_{0}d\Omega\big[\Omega^{\dag}_{j,l^{\prime},j_z}\;i{\sigma}\cdot(\mathbf n\times\bar{\theta})\;\Omega_{j,l,j_z^{\prime}}\big],\nonumber\\
\end{eqnarray}
where $\sigma=(\sigma_{i})$ are the Pauli matrices; $\mathbf r=r\mathbf n$, $\mathbf n$ is a unit vector in a
direction of $\mathbf r$. Note that for $2S_{1/2}, 2P_{1/2}$ and $2P_{3/2}$ cases, $g(r)$ and $f(r)$ are real functions, thus we can simplify
(\ref {E bar theta}) as
\begin{eqnarray}\label{E bar theta simple}
E^{\bar{\theta}}_{j_zj_z^{\prime}}(nL_j)=\frac{1}{4\alpha}\bar{\rho}(nL_j)\bar{\Theta}_{j_zj_z^{\prime}}(nL_j),
\end{eqnarray}
with
\begin{eqnarray}\label{bar rho and bar theta}
\bar{\rho}(nL_j)&=&\int^{\infty}_{0}drr^{3}g(r)\;f(r), \nonumber\\
\bar{\Theta}_{j_zj_z^{\prime}}(nL_j)&=&\int^{4\pi}_{0}d\Omega\big[\Omega^{\dag}_{j,l^{\prime},j_z}\;i{\sigma}\cdot(\mathbf n\times\bar{\theta})\;\Omega_{j,l,j_z^{\prime}}\big]\\
&&-\int^{4\pi}_{0}d\Omega\big[\Omega^{\dag}_{j,l,j_z}\;i{\sigma}\cdot(\mathbf n\times\bar{\theta})\;\Omega_{j,l^{\prime},j_z^{\prime}}\big]\;.\nonumber
\end{eqnarray}
In the following sections we will calculate the
$\theta$ and $\bar{\theta}$ modifications of the $2S$ and $2P$ energy
levels. In our calculation, we choose $\theta_1=\theta_2=0$, $\theta_3=\theta$  as well as $\bar{\theta}_1=\bar{\theta}_2=0$ and $\bar{\theta}_3=\bar{\theta}$, that can be achieved by rotational invariance or redefinition of coordinates.

\subsection{Relativistic NCPS correction of $2S_{1/2}$ and  $2P_{1/2}$}
In this subsection we calculate the $\theta$ and the $\bar\theta$ corrections for  the energy levels $2S_{1/2}$ and  $2P_{1/2}$.
From (\ref{theta correction 2 }) the $\theta$ correction for the level $2S_{1/2}$ ($n=2, n^{\prime}=1, j=1/2, l=0, j=l+1/2, l^{\prime}=1, j_z=\pm1/2$) is
\begin{eqnarray}\label{E theta}
E^{\theta}_{j_zj_z^{\prime}}(2S_{1/2})=-\frac{Ze^{2}}{4\alpha^3}\rho(2S_{1/2})\Theta_{j_zj_z^{\prime}}(2S_{1/2}),
\end{eqnarray}
where the $\rho(2S_{1/2})$ and the $\Theta_{j_zj_z^{\prime}}(2S_{1/2})$ are follows from (\ref{rho and theta}) and by some calculation we obtain
\begin{eqnarray}
\Theta_{j_zj_z^{\prime}}(2S_{1/2})=\int^{4\pi}_{0}d\Omega\big[\Omega^{\dag}_{1/2,0,j_z}(\mathbf
L\cdot \theta)\Omega_{1/2,0,j_z^{\prime}}\big]=0\;.\nonumber
\end{eqnarray}
Thus the $\theta$ correction for the relativistic energy level $2S_{1/2}$ is
\begin{equation}\label{E theta 2S1/2 final}
E^{\theta}_{j_zj_z^{\prime}}(2S_{1/2})=0\;.
\end{equation}
It is clear that the $H^\theta$ does not modify  the energy level  $2S_{1/2}$ .

From (\ref{E bar theta simple}) the $\bar{\theta}$ correction for the $2S_{1/2}$ level is
\begin{eqnarray}
E^{\bar{\theta}}_{j_zj_z^{\prime}}(2S_{1/2})=\frac{1}{4\alpha}\bar{\rho}(2S_{1/2})\bar{\Theta}_{j_zj_z^{\prime}}(2S_{1/2}),
\end{eqnarray}
here the $\bar{\rho}(2S_{1/2})$ and the $\bar{\Theta}_{j_zj_z^{\prime}}(2S_{1/2})$ are follows from (\ref{bar rho and bar theta}):
\begin{eqnarray}\label{delta E theta bar}
\bar{\rho}(2S_{1/2})&=&\int^{\infty}_{0}drr^{3}g(r)\;f(r) \nonumber\\
&=&\frac{1}{8m\eta_1}\big[\frac{\beta^2_1}{\eta_1+1}-\beta^2_1(\eta_1+1)
+2\beta_1(\beta_1+1)\nonumber\\
&&\times(\eta_1+1)-(\beta_1+1)(\beta_1+2)(\eta_1+1)\big]\;,\nonumber\\
\end{eqnarray}
with
\begin{eqnarray}\label{gamma 1 and beta1}
\gamma_1=\sqrt{1-(Ze^2)^2} \;,\;\; \beta_1=2\gamma_1+1,\;\; \eta_1=\frac{(1+\gamma_1)m}{E_{n,j}}.\nonumber\\
\end{eqnarray}
\begin{eqnarray}\
\bar{\Theta}_{j_zj_z^{\prime}}(2S_{1/2})&=&\int^{4\pi}_{0}d\Omega\big[\Omega^{\dag}_{1/2,1,j_z}\;i{\sigma}\cdot(\mathbf n\times\bar{\theta})\;\Omega_{1/2,0,j_z^{\prime}}\big]\nonumber\\
&-&\int^{4\pi}_{0}d\Omega\big[\Omega^{\dag}_{1/2,0,j_z}\;i{\sigma}\cdot(\mathbf n\times\bar{\theta})\;\Omega_{1/2,1,j_z^{\prime}}\big].\nonumber\\
\end{eqnarray}
Thus the matrix $\bar{\Theta}(2S_{1/2})$ is
\begin{eqnarray}
\bar{\Theta}(2S_{1/2})=\frac{4}{3}\left(
                               \begin{array}{cc}
                                 \bar{\theta}_{3} & 0 \\
                                0 & \bar{\theta}_{3} \\
                               \end{array}
                             \right).
\end{eqnarray}
Corresponding eigenvalues of the matrix $\bar{\Theta}(2S_{1/2})$ are
\begin{equation}
\bar{\Lambda}_{\pm1/2}(2S_{1/2})=\frac{4}{3}\mid \bar{\theta}\mid\;,
\end{equation}
where $|{\bar{\theta}}|=\sqrt{\bar{\theta_{i}}\bar{\theta_{i}}}$ . The $\bar{\theta}$ correction to the relativistic energy level
$2S_{1/2}$ is
\begin{eqnarray}\label{E bar theta 2S1/2 final}
E^{\bar{\theta}}_{\pm1/2}(2S_{1/2})&=&\frac{1}{4\alpha}\bar{\rho}(2S_{1/2})\bar{\Lambda}_{\pm1/2}(2S_{1/2})\nonumber\\
&=&\frac{1}{24\alpha m \eta_1}\mid \bar{\theta}\mid \big[\frac{\beta^2_1}{\eta_1+1}-\beta^2_1(\eta_1+1)\nonumber\\
&&+2\beta_1(\beta_1+1)(\eta_1+1)-(\beta_1+1)\nonumber\\
&&\times(\beta_1+2)(\eta_1+1)\big]\;.
\end{eqnarray}
Note that the $\bar{\theta}$ correction shifts the relativistic energy level $2S_{1/2}$.
From (\ref{Delta E}), (\ref{E theta 2S1/2 final}) and (\ref{E bar theta 2S1/2 final}) the relativistic NCPS correction for the energy level $2S_{1/2}$ is
\begin{equation}\label{NCPS for 2S1/2}
\Delta E_{\pm1/2}(2S_{1/2})=\Delta E_{2,1/2}+E^{\bar{\theta}}_{\pm1/2}(2S_{1/2}).
\end{equation}

In the following we calculate the $\theta$ and the $\bar\theta$ corrections for the energy level $2P_{1/2}$.
From (\ref{theta correction 2 }) the $\theta$ correction for the level $2P_{1/2}$ ($n=2, n^{\prime}=1, j=1/2, l=1, j=l-1/2, l^{\prime}=0 ,j_z=\pm1/2$) is
\begin{equation}
E^{\theta}_{j_z,j_z^\prime}(2P_{1/2})=-\frac{Ze^{2}}{4{\alpha}^3}\rho(2P_{1/2})\Theta_{j_z,j_z^\prime}(2P_{1/2}),
\end{equation}
where the radial integral $\rho(2P_{1/2})$ and the matrix elements $\Theta_{j_z,j_z^\prime}(2P_{1/2})$ can obtained from  (\ref{rho and theta})
\begin{eqnarray}
\rho(2P_{1/2})&=&\int^{\infty}_{0}\frac{dr}{r}|g(r)|^2\nonumber\\
&=&\frac{2\lambda_1^{3}(\lambda^{+}_1)^{2}}{\eta_1(\beta_1-1)(\beta_1-2)(\beta_1-3)}\nonumber\\
&& \times \big\{\beta_1\frac{(\eta_1-2)^2}{\eta_1-1}-2(\eta_1-2)(\beta_1-3)\nonumber\\
&&+\big(\frac{\eta_1-1}{\beta_1}\big)(\beta_1-2)(\beta_1-3)\big\}\;,
\end{eqnarray}
here $\lambda_1$ , $\lambda^{+}_1$ are given in (\ref {def of eta
gamma}) and $\beta_1$ , $\eta_1$ are given in (\ref{gamma 1 and beta1}).
From (\ref{thetamm}) we have
\begin{eqnarray}
\Theta_{j_zj_z^{\prime}}(2P_{1/2})=\int^{4\pi}_{0}d\Omega\big[\Omega^{\dag}_{1/2,1,j_z}(\mathbf
L\cdot \theta)\Omega_{1/2,1,j_z^{\prime}}\big]\;.
\end{eqnarray}
By some calculation we get the following matrix
\begin{eqnarray}
\Theta(2P_{1/2})=\frac{2}{3}\left(
             \begin{array}{cc}
               \theta_{3}& \theta_{-}\\
               \theta_{+} & -\theta_{3} \\
             \end{array}
           \right),
\end{eqnarray}
where $\theta_{\pm}=\theta_{1}\pm i\theta_{2}$. The eigenvalues of the matrix $\Theta(2P_{1/2})$ are
\begin{equation}
\Lambda_{\pm1/2}(2P_{1/2})=\pm\frac{2}{3}|{\mathbf\theta}|\;,
\end{equation}
here $|{\mathbf\theta}|=\sqrt{\theta_{i}\theta_{i}}$ . Then the $\theta$ correction to the relativistic energy of the $2P_{1/2}$
level is
\begin{eqnarray}\label{E theta for 2P1/2}
E^{\theta}_{\pm1/2}(2P_{1/2})&=&-\frac{Ze^{2}}{4{\alpha}^3}\rho(2P_{1/2})\Lambda_{\pm1/2}(2P_{1/2})\nonumber\\
&=&\mp\frac{Ze^{2}}{3{\alpha}^3}|{\mathbf\theta}|\bigg\{\frac{\lambda_1^{3}(\lambda^{+}_1)^{2}}{\eta_1(\beta_1-1)(\beta_1-2)(\beta_1-3)}\nonumber\\
&& \times \big[\beta_1\frac{(\eta_1-2)^2}{\eta_1-1}-2(\eta_1-2)(\beta_1-3)\nonumber\\
&&+\big(\frac{\eta_1-1}{\beta_1}\big)(\beta_1-2)(\beta_1-3)\big]\bigg\}\;.
\end{eqnarray}
We can see that the degenerate level $2P_{1/2}$ splits into two sublevels by the $\theta$ correction.

From (\ref{E bar theta simple}) the $\bar{\theta}$ correction for the $2P_{1/2}$ level is
\begin{eqnarray}
E^{\bar{\theta}}_{j_zj_z^{\prime}}(2P_{1/2})=\frac{1}{4\alpha}\bar{\rho}(2P_{1/2})\bar{\Theta}_{j_zj_z^{\prime}}(2P_{1/2}),
\end{eqnarray}
here the $\bar{\rho}(2P_{1/2})$ and the $\bar{\Theta}_{j_zj_z^{\prime}}(2P_{1/2})$ are follows from (\ref{bar rho and bar theta}):
\begin{eqnarray}
\bar{\rho}(2P_{1/2})&=&\int^{\infty}_{0}drr^{3}g(r)\;f(r)\nonumber\\
&=&\frac{1}{8m\eta_1}\big[\frac{\beta^2_1}{\eta_1-1}-\beta^2_1(\eta_1-1)+2\beta_1(\beta_1+1)\nonumber\\
&&\times(\eta_1-1)-(\beta_1+1)(\beta_1+2)(\eta_1-1)\big]\;,\nonumber\\
\end{eqnarray}
with $\beta_1$ and $\eta_1$ are given in (\ref{gamma 1 and beta1}).
\begin{eqnarray}\
\bar{\Theta}_{j_zj_z^{\prime}}(2P_{1/2})&=&\int^{4\pi}_{0}d\Omega\big[\Omega^{\dag}_{1/2,0,j_z}\;i{\sigma}\cdot(\mathbf n\times\bar{\theta})\;\Omega_{1/2,1,j_z^{\prime}}\big]\nonumber\\
&-&\int^{4\pi}_{0}d\Omega\big[\Omega^{\dag}_{1/2,1,j_z}\;i{\sigma}\cdot(\mathbf n\times\bar{\theta})\;\Omega_{1/2,0,j_z^{\prime}}\big].\nonumber\\
\end{eqnarray}
Again by some calculation we obtain the matrix $\bar{\Theta}(2P_{1/2})$
\begin{eqnarray}
\bar{\Theta}(2P_{1/2})=-\frac{4}{3}\left(
                               \begin{array}{cc}
                                 \bar{\theta}_{3} & 0 \\
                                0 & \bar{\theta}_{3} \\
                               \end{array}
                             \right),
\end{eqnarray}
the eigenvalues of $\bar{\Theta}(2P_{1/2})$ are
\begin{equation}
\bar{\Lambda}_{\pm1/2}(2P_{1/2})=-\frac{4}{3}\mid\bar{\theta}\mid\;.
\end{equation}
Therefore the $\bar{\theta}$ correction to the relativistic energy
of the $2P_{1/2}$ level is
\begin{eqnarray}\label{delta E theta bar for 2P1/2}
E^{\bar{\theta}}_{\pm1/2}(2P_{1/2})&=&\frac{1}{4\alpha}\bar{\rho}(2P_{1/2})\bar{\Lambda}_{\pm1/2}(2P_{1/2})\nonumber\\
&=& -\frac{1}{24\alpha m \eta_1}\mid\bar{\theta}\mid \big[\frac{\beta^2_1}{\eta_1-1}-\beta^2_1(\eta_1-1)\nonumber\\
&&+2\beta_1(\beta_1+1)(\eta_1-1)-(\beta_1+1)\nonumber\\
&&\times(\beta_1+2)(\eta_1-1)\big] \;.
\end{eqnarray}
The $\bar{\theta}$ correction shifts the energy level $2P_{1/2}$.
Therefor the relativistic NCPS energy correction for the level $2P_{1/2}$ is
\begin{eqnarray}\label{NCPS for 2P1/2}
\Delta  E_{\pm1/2}(2P_{1/2})&=&\Delta E_{2,1/2}+E_{\pm1/2}^{\theta}(2P_{1/2})\nonumber\\
&&+E^{\bar{\theta}}_{\pm1/2}(2P_{1/2}).
\end{eqnarray}
From (\ref{E theta 2S1/2 final}),(\ref{E bar theta 2S1/2 final}), (\ref{NCPS for 2S1/2}), (\ref{E theta for 2P1/2}), (\ref{delta E theta bar for 2P1/2}) and (\ref{NCPS for 2P1/2}) one can see that the $\theta$ correction splits the  original energy levels $2S_{1/2}$, $2P_{1/2}$  into three sublevels $2S_{1/2}$ and $2P^{\pm1/2}_{1/2}$, and the $\bar{\theta}$ correction shifts the energy levels. Modifications of these energy levels are illustrated in Fig.1 for $Z=1$ (hydrogen atom).
\subsection{Relativistic NCPS correction for the $2P_{3/2}$ level}
In this subsection we calculate the $\theta$ and $\bar\theta$ corrections for  the energy
level  $2P_{3/2}$.
For $2P_{3/2}$ level ($n=2, n^{\prime}=0, j=3/2, l=1, j=l+1/2, l^{\prime}=2, j_z=\pm1/2,\pm3/2$),
 from (\ref{theta correction 2 }) the $\theta$ correction is
\begin{eqnarray}
E^{\theta}_{j_zj_z^{\prime}}(2P_{3/2})=-\frac{Ze^{2}}{4\alpha^3}\rho(2P_{3/2})\Theta_{j_zj_z^{\prime}}(2P_{3/2}),
\end{eqnarray}
here the $\rho(2P_{3/2})$ and the $\Theta_{j_zj_z^{\prime}}(2P_{3/2})$ have the following forms by (\ref{rho and theta})
\begin{eqnarray}
\rho (2P_{3/2})&=&\int^{\infty}_{0}\frac{dr}{r}|g(r)|^2 \nonumber\\ &=&\frac{4\lambda_{0}^{3}(\lambda^{+}_0)^{2}}{(\beta_0-1)(\beta_0-2)(\beta_0-3)}\;,\nonumber\\
\end{eqnarray}
where $\lambda_0$ and $\lambda^{+}_0$ are given in (\ref {def of eta
gamma}), and
\begin{equation}\label{gamma 0 and beta 0}
\gamma_0=\sqrt{4-(Ze^2)^2} \;,\;\;\; \beta_0=2\gamma_0+1\;.
\end{equation}
The matrix elements $\Theta_{j_zj_z^{\prime}}(2P_{3/2})$  are given by
\begin{eqnarray}
\Theta_{j_zj_z^{\prime}}(2P_{3/2})=\int^{4\pi}_{0}d\Omega\big[\Omega^{\dag}_{3/2,1,j_z}(\mathbf
L\cdot\theta)\Omega_{3/2,1,j_z^{\prime}}\big]\;,
\end{eqnarray}
 the corresponding matrix $\Theta(2P_{3/2})$ is
\begin{eqnarray}
\Theta(2P_{3/2})=\frac{1}{3}\left(
             \begin{array}{cccc}
               -3\theta_{3} & \sqrt{3}\theta_{+} & 0 & 0 \\
                \sqrt{3}\theta_{-}& -\theta_{3} & 2\theta_{+}& 0 \\
               0 & 2\theta_{-}& \theta_{3} & \sqrt{3}\theta_{+} \\
               0 & 0 & \sqrt{3}\theta_{-} & 3\theta_{3}\\
             \end{array}\;
           \right).
\end{eqnarray}
Corresponding four nondegenerate eigenvalues $\Lambda_{\pm3/2}(2P_{3/2})$ and
$\Lambda_{\pm1/2}(2P_{3/2})$ are
\begin{eqnarray}
\Lambda_{\pm3/2}(2P_{3/2})=\pm|{\mathbf\theta}|\;,\hspace{0.5cm}\Lambda_{\pm1/2}(2P_{3/2})=\pm\frac{1}{3}|\theta|\;,
\end{eqnarray}
and hence, the $\theta$ correction for the $2P_{3/2}$ level have the following form
\begin{eqnarray}\label{E theta to 2P3/2}
E^{\theta}_{\pm3/2}(2P_{3/2})&=&\mp\frac{Z{e}^{2}}{\alpha^3}|\theta| \bigg\{\frac{\lambda_{0}^{3}(\lambda^{+}_0)^{2}}{(\beta_0-1)(\beta_0-2)(\beta_0-3)}\bigg\}\;,\nonumber\\
E^{\theta}_{\pm1/2}(2P_{3/2})&=&\frac{1}{3}E^{\theta}_{\pm3/2}(2P_{3/2})\;.
\end{eqnarray}
Note that the $\theta$ correction splits the $2P_{3/2}$ level into four nondegenerate sublevels $2P^{\pm3/2}_{3/2}$ and
$2P^{\pm1/2}_{3/2}$.

From (\ref{E bar theta simple}) the $\bar{\theta}$ correction for level $2P_{3/2}$ is
\begin{eqnarray}
E^{\bar{\theta}}_{j_zj_z^{\prime}}(2P_{3/2})=\frac{1}{4\alpha}\bar{\rho}(2P_{3/2})\bar{\Theta}_{j_zj_z^{\prime}}(2P_{3/2})\;,
\end{eqnarray}
where the $\bar{\rho}(2P_{1/2})$ and  $\bar{\Theta}_{j_zj_z^{\prime}}(2P_{1/2})$ are obtained from  (\ref{bar rho and bar theta})
\begin{eqnarray}\label{delta E theta bar}
\bar{\rho}(2P_{3/2})=\int^{\infty}_{0}drr^{3}g(r)\;f(r)=-\frac{\beta_0(\eta_0+2)}{8m\eta_0}\;,
\end{eqnarray}
here $\beta_0$ is given in (\ref{gamma 0 and beta 0}) and
\begin{equation}
\eta_0=\frac{m\gamma_0}{E_{n,j}}\;,\;\;\;\;\gamma_0=\sqrt{4-(Ze^2)^2}\;.
\end{equation}
\begin{eqnarray}
\bar{\Theta}_{j_zj_z^{\prime}}(2P_{3/2})
&=&\int^{4\pi}_{0}d\Omega\big[\Omega^{\dag}_{3/2,2,j_z}\;i{\sigma}\cdot(\mathbf n\times\bar{\theta})\;\Omega_{3/2,1,j_z^{\prime}}\big]\nonumber\\
&-&\int^{4\pi}_{0}d\Omega\big[\Omega^{\dag}_{3/2,1,j_z}\;i{\sigma}\cdot(\mathbf n\times\bar{\theta})\;\Omega_{3/2,2,j_z^{\prime}}\big]\;.\nonumber\\
\end{eqnarray}
Then the matrix $\bar\Theta(2P_{3/2})$ is
\begin{eqnarray}
\bar{\Theta}(2P_{3/2})=\frac{2}{5}\left(
             \begin{array}{cccc}
               4\bar{\theta}_{3} & -\frac{1}{\sqrt{3}}\bar{\theta}_{+} & 0 & 0 \\
                -\frac{1}{\sqrt{3}}\bar{\theta}_{-}& \frac{8}{3}\bar{\theta}_{3} & 0 & 0 \\
               0 & 0 & \frac{8}{3}\bar{\theta}_{3} &\frac{1}{\sqrt{3}}\bar{\theta}_{+} \\
               0 & 0 & \frac{1}{\sqrt{3}}\bar{\theta}_{-} & 4\theta_{3}\\
             \end{array}\nonumber
           \right),
\end{eqnarray}
where $\bar{\theta}_{\pm}=\bar{\theta}_{1}\pm i\bar{\theta}_{2}$.
The eigenvalues of the matrix $\bar{\Theta}(2P_{3/2})$ are
\begin{eqnarray}
\bar{\Lambda}_{\pm3/2}(2P_{3/2})=\frac{8}{5}|\bar{\theta}|\;,\hspace{0.5cm} \bar{\Lambda}_{\pm1/2}(2P_{3/2})=\frac{16}{15}|\bar{\theta}|\;.
\end{eqnarray}
The $\bar{\theta}$ correction for the relativistic energy level $2P_{3/2}$ are
\begin{eqnarray}\label{E bar theta correction to 2p3/2}
E^{\bar{\theta}}_{\pm3/2}(2P_{3/2})=-\frac{\beta_0(\eta_0+2)}{20\alpha m\eta_0}|\bar{\theta}|,\nonumber\\
E^{\bar{\theta}}_{\pm1/2}(2P_{3/2})=\frac{2}{3}E^{\bar{\theta}}_{\pm3/2}(2P_{3/2}).
\end{eqnarray}
Therefor the $\bar{\theta}$ correction  shifts the energy levels.
Finally, the relativistic NCPS energy corrections for the level $2P_{3/2}$ are
\begin{eqnarray}\label{NCPS for 2P3/2}
\Delta \hat E_{\pm3/2}(2P_{3/2})&=&\Delta E_{2,3/2}+E_{\pm3/2}^{\theta}(2P_{3/2})\nonumber\\
&&+E^{\bar{\theta}}_{\pm3/2}(2P_{3/2}),\nonumber\\
\Delta \hat E_{\pm1/2}(2P_{3/2})&=&\Delta E_{2,3/2}+E_{\pm1/2}^{\theta}(2P_{3/2})\nonumber\\
&&+E^{\bar{\theta}}_{\pm1/2}(2P_{3/2}).
\end{eqnarray}
From (\ref{E theta to 2P3/2}), (\ref{E bar theta correction to 2p3/2}) and (\ref{NCPS for 2P3/2})one can see that the relativistic energy level $2P_{3/2}$ is split into four nondegenerate sublevels by the $\theta$ correction and shifted by the $\bar{\theta}$ correction.

The energy levels and their modified energy levels of the hydrogen atom due to space-space, and momentum-momentum noncommutativity for the relativistic case are shown in Fig.1.
The subenergy level spacings for the relativistic energy levels $2S_{1/2}$ and $2P_{1/2}$ are as follows (in units of $eV/m^2$)
\begin{eqnarray}
\Delta  E(2P^{-1/2}_{1/2}\rightarrow2S_{1/2}) &=&6.75\times10^{19}\frac{|\theta|}{\alpha^3}+8.38\times10^6\frac{|\bar{\theta|}}{\alpha},\nonumber\\
\Delta  E(2S_{1/2}\rightarrow2P^{1/2}_{1/2}) &=&6.75\times10^{19}\frac{|\theta|}{\alpha^3}-8.38\times10^6\frac{|\bar{\theta|}}{\alpha}.\nonumber
\end{eqnarray}
The subenergy level spacings for the relativistic energy level $2P_{3/2}$ are as follows (in units of $eV/m^2$)
\begin{eqnarray}
\Delta  E(2P^{-3/2}_{3/2}\rightarrow 2P^{-1/2}_{3/2})&=&6.75\times10^{19}\frac{|\theta|}{\alpha^3}-8.38\times10^6\frac{|\bar{\theta|}}{\alpha},\nonumber\\
\Delta E(2P^{-1/2}_{3/2}\rightarrow 2P^{1/2}_{3/2})&=&6.75\times10^{19}\frac{|\theta|}{\alpha^3},\nonumber\\
\Delta  E(2P^{1/2}_{3/2}\rightarrow 2P^{3/2}_{3/2})&=&6.75\times10^{19}\frac{|\theta|}{\alpha^3}+8.38\times10^6\frac{|\bar{\theta|}}{\alpha}.\nonumber
\end{eqnarray}

  \begin{figure}
         \begin{minipage}[]{85mm}
          \includegraphics[width=8.6cm]{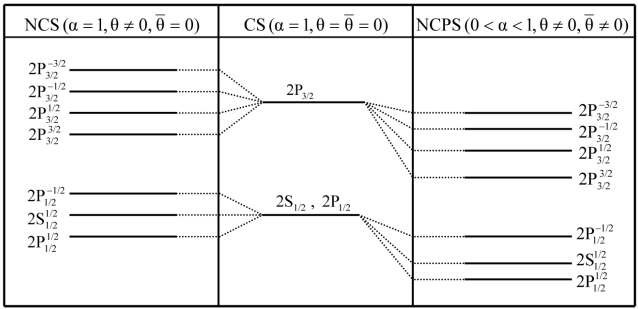}
   \caption{Modifications for relativistic energy levels of hydrogen atom on a NCPS}
   \end{minipage}
     \label{Fig1}
  \end{figure}

\section{conclusion}
Since the noncommutative parameters $\theta$ and $\bar\theta$ is very small compared to the length scales of the system, one can always treat the noncommutative effects as some perturbation of the commutative counter part. Thus we calculated the  energy levels $2S_{1/2}$, $2P_{1/2}$ and $2P_{3/2}$ of the relativist hydrogen-like atom on a NCPS by using perturbation method. We found that the space-space noncommutativity  splits the degenerate energy levels $2S_{1/2}$, $2P_{1/2}$  into three sublevels $2S_{1/2}$ and $2P^{\pm1/2}_{1/2}$,
 as well as splits the level $2P_{3/2}$  into four nondegenerate sublevels $2P^{\pm1/2}_{3/2}$ and $2P^{\pm3/2}_{3/2}$, such that new transition channels are allowed between subenergy levels. We also show that the momentum-momentum noncommutativity  shifts the energy levels. The $\theta$ and the $\bar\theta$ modification of the energy levels for the relativistic case are shown in Fig.1.

To impose some bounds on
the value of the noncommutativity parameters $\theta$, $\bar{\theta}$, one needs
experimental data with a high accuracy.

If $\bar\theta=0$, it leads $\alpha=1$, then $\Delta E_{n,j}=E^{\bar\theta}=0$, we get the NCS results of Ref.\cite{HAncs}. If $\theta=\bar \theta=0$, then $\Delta E_{n,j}=E^{\theta}=E^{\bar{\theta}}=0$, we obtain the usual relativistic quantum mechanics results of Refs.\cite{Greiner}-\cite{Akhiezer}.
The results in this paper suggest that high precision
measurements in quantum mechanical systems may be able to reveal the
noncommutativity of space and phase space.

\section{Acknowledgments}
H. Masum is grateful to Kai Ma for many useful discussions. This work is supported by the National Natural Science Foundation of China (10965006 and 11165014).


\begin{thebibliography}{*}
\bibitem{NCQEDGG}S. Godfrey and M. A. Doncheski, Phys. Rev. \textbf{D65}, (2001) 015005.

\bibitem{NCPM}M. Haghighat and M. M. Ettefaghi, Phys. Rev.  \textbf{D70} (2004) 034017.

\bibitem{NCEPGGG}A. Devoto, S. DiChiara, and W. W. Repko, Phys. Rev. \textbf{D72},
(2005) 056006.

\bibitem{NCQED}X. Calmet, Eur. Phys. J. \textbf{C50}, (2007) 113.

\bibitem{10}Chaichian M, Presnajder P, Sheikh-Jabbari M M and Tureanu A 2002 Phys. Lett. B \textbf{527} 149-54
\bibitem{11}H. Falomir, J. Gamboa, M. Loewe, F. M\'{e}ndez, J. C. Rojas, Phys. Rev. \textbf{D66}, (2002) 045018.

\bibitem{12} K. Li, S. Dulat,  Eur. Phys. J. C \textbf{46}, 825 (2006).

\bibitem{13} B. Mirza  and M. Zarei, Eur. Phys. J. C \textbf{32} 583,  2004.

\bibitem{14}K. Li, J.-H. Wang, Eur. Phys. J. \textbf{C50}, (2007) 1007.

\bibitem{15}B. Mirza, R. Narimani, M. Zarei, Eur. Phys. J. \textbf{C48}, (2006)641;

\bibitem{16} S. Dulat, K. Li,  Eur. Phys. J. C \textbf{54}, 333 (2008).

\bibitem{17}B. Harms and O. Micu, J. Phys. \textbf{A40}, (2007) 10337.

\bibitem{18}O. F. Dayi and A. Jellal, J. Math. Phys. \textbf{43}, (2002) 4592;
J. Math. Phys. \textbf{45}, (2004) 827(E);
A. Kokado, T. Okamura, and T. Saito, Prog. Theor. Phys. \textbf{110}, (2003) 975;
S. Dulat and K. Li, Eur.Phys. J. \textbf{C60}, (2009) 163.

\bibitem{19}B. Chakraborty, S. Gangopadhyay, and A. Saha,
Phys. Rev. \textbf{D70}, (2004) 107707;
F. G. Scholtz, B. Chakraborty, S. Gangopadhyay, and A. G. Hazra,
Phys. Rev.  \textbf{D71}, (2005) 085005;
F. G. Scholtz, B.Chakraborty, S. Gangopadhyay, and J. Govaerts,
J. Phys.\textbf{A38}, (2005) 9849.

\bibitem{20}O. F. Dayi and M. Elbistan, Phys. Lett.  \textbf{A373},
(2009) 131.

\bibitem{21} Kai Ma, Sayipjamal Dulat,  Phys.Rev. \textbf{A84} (2011) 012104


\bibitem{HAncs}M. Chaichian, M. M. Sheikh-Jabbari, and A. Tureanu, Phys. Rev. Lett. \textbf{86}, (2001) 2716.

 \bibitem{HAncps} Li Kang, CHAMOUN Nidal, Chin. Phys. Lett. \textbf{ 23}, 5(2006).



\bibitem{Haghighat} M. Haghighat and F. Loran, Phys. Rev.  \textbf{D67} (2003) 096003.

\bibitem{NCPR}J.-Z. Zhang, Phys. Rev. Lett.  \textbf{93} (2004) 043002.

\bibitem{RHAncs}T. C. Adorno, M. C. Baldiotti, M. Chaichian, D. M. Gitman, and A.
Tureanu, Phys. Lett. \textbf{B682}, (2009) 235.


\bibitem{Greiner} W. Greiner, Relativisic Quantum Mechanics: Wave Equation (Springer, 2000)

\bibitem {Bethe} H. A. Bethe, Edwin E. Salpeter, Quantum Mechanics of One- and Two-Electron Atoms, Springer-Verlag, 1957.

\bibitem {Rose}  M. E. Rose, Relativistic Electron Theory, John Wiley and Sons, 1961.

\bibitem  {Akhiezer} A. I. Akhiezer, V.B. Berestetskii, Quantum Electrodynamics, Interscience Publishers, 1965.

\bibitem {Voronov}  B. L. Voronov, D. M. Gitman and I. V. Tyutin (The Dirac Hamiltonian with a superstrong Coulomb field), Theoretical and Mathematical Physics , 150, 34 (2007)

\bibitem{hypergeometric confluent and Gamma} I.S. Gradshteyn, I.M. Ryzhik, Table of Integrals, Series, and Products,
7th ed., Academic Press, 2007.

\end{thebibliography}
\end{document}